\documentclass[aip,reprint]{revtex4-2}
\usepackage{upgreek}
\usepackage{graphicx}
\usepackage{bm}
\usepackage{amsmath, amsthm, amssymb}
\usepackage{latexsym}
\usepackage{dcolumn}

\begin{document}
	
\title{Frequency limits of sequential readout for sensing AC magnetic fields using nitrogen-vacancy centers in diamond}

\author{Santosh \surname{Ghimire}}
\affiliation{Korea Research Institute of Standards and Science, Daejeon 34113, Republic of Korea}
\author{Seong-Joo \surname{Lee}}
\affiliation{Korea Research Institute of Standards and Science, Daejeon 34113, Republic of Korea}
\author{Sangwon \surname{Oh}}
\affiliation{Korea Research Institute of Standards and Science, Daejeon 34113, Republic of Korea}
\author{Jeong Hyun \surname{Shim}}
\email{jhshim@kriss.re.kr}
\affiliation{Korea Research Institute of Standards and Science, Daejeon 34113, Republic of Korea}
\affiliation{Department of Applied Measurement Science, University of Science and Technology, Daejeon 34113, Republic of Korea}
	
\date{\today}
	\begin{abstract}
The nitrogen-vacancy (NV) centers in diamond have ability to sense alternating-current (AC) magnetic fields with high spatial resolution. However, the frequency range of AC sensing protocols based on dynamical decoupling (DD) sequences has not been thoroughly explored experimentally. In this work, we aimed to determine the sensitivity of ac magnetic field as a function of frequency using sequential readout method. The upper limit at high frequency is clearly determined by Rabi frequency, in line with the expected effect of finite DD-pulse width. In contrast, the lower frequency limit is primarily governed by the duration of optical repolarization rather than the decoherence time (T$_2$) of NV spins. This becomes particularly crucial when the repetition (dwell) time of the sequential readout is fixed to maintain the acquisition bandwidth. The equation we provide successfully describes the tendency in the frequency dependence. In addition, at the near-optimal frequency of 1 MHz, we reached a maximum sensitivity of 229 pT/$\sqrt{\mathrm{Hz}}$ by employing the XY4-(4) DD sequence.
	\end{abstract}

	\maketitle

\section{Introduction}
Sensing alternating-current (AC) magnetic fields in the range of kilohertz to megahertz has been used in various applications, including nuclear magnetic resonance (NMR),\cite{Glenn2018} magnetic induction tomography,\cite{Deans2021} and magnetic communications.\cite{Gerginov2017} Solid-state quantum spins, such as negatively charged nitrogen-vacancy (NV) centers in diamond\cite{Balasubramanian2008, Taylor2008,Wolf2015}, are among the promising sensors for ac magnetic fields due to their high spatial resolution, which ranges from micro to nanometer.\cite{Balasubramanian2008,Taylor2008,Maze2008,Degen2008,Balasubramanian2009,Maurer2010,
		Dolde2011,Grinolds2011,Wolf2015,Appel2015,Siyushev2019,Barson2021}  The controllable density and depth of the quantum spins near the surface of host materials provide a reduced standoff distance from sources, resulting in  such high resolution. In addition, the protocols for sensing ac magnetic fields allow control over both sensible frequency and bandwidth.\cite{Degen2017} Moreover, the bandwidth that NV spins provide is often significantly wider than that of radio-frequency optically pumped magnetometers.\cite{Savukov2005} The boardband sensing using solid-state spin sensors enables the high-fidelity reception of rapidly varying or modulated ac signals.
	
	AC-field sensing protocols called dynamical decoupling (DD) exploit a multiple number of phase-refocusing  $\pi$ pulses.\cite{Rondin2014,Degen2017,Suter2017} The stroboscopic pulse train periodically flips the quantum states and hence filters out a specific frequency, reducing the frequency window and detuning the quantum spins from surrounding magnetic noises. This increases the coherence time T$_2$ of quantum spins \cite{deLange2011,Pham2011,Pham2012} and thus enhances sensitivity to AC magnetic fields.\cite{Taylor2008,Acosta2009,Edlyn2019} However, when detecting an oscillating signal with a time-varying or modulated envelope, such as nuclear spin precessions in NMR, the signal often persists longer than the T$_2$ duration. To address this issue, sequential readout (SR) scheme or quantum heterodyne (Qdyne) has been developed, based on the repetition of a block consisting of a DD sequence and projective readout. The SR method works because the phase of the AC magnetic field continues even though the quantum states collapse during the projective readout.\cite{Jordan2017,Boss2017,Schmitt2017} In principle, the spectral resolution achievable with SR is limited only by the stability of external clocks synchronizing instrumentations, regardless of T$_2$.\cite{Boss2017} This feature enabled sensing of an AC magnetic field with sub-millihertz resolution\cite{Schmitt2017} and acquiring of high-resolution NMR spectra from micron-scale liquid samples.\cite{Glenn2018, Smits2019, Bruckmaier2023}
	
	The range of frequency to which the SR method is sensitive is certainly crucial for its applications in the field. As the sensing frequency increases, the time interval $\tau$ in Fig. 1 (a) decreases, and eventually the interrogation window closes. Considering the finite width of the $\pi$ pulses, one can expect Rabi-oscillation frequency to be the high-frequency limit. The low-frequency limit may be determined by  decoherence process, as the maximum $\tau$ will be imposed by the value of T$_2$. These expectations, however, have not been experimentally investigated yet. In this study, we investigate the frequency dependence of AC-field sensitivity using an ensemble of NV centers in diamond. By accounting for the finite widths of $\pi$ pulses and the duration of  optical readout/repolarization, we successfully explain the obtained AC-field sensitivity.

\begin{figure}[t!]
	\includegraphics[width=1\columnwidth]{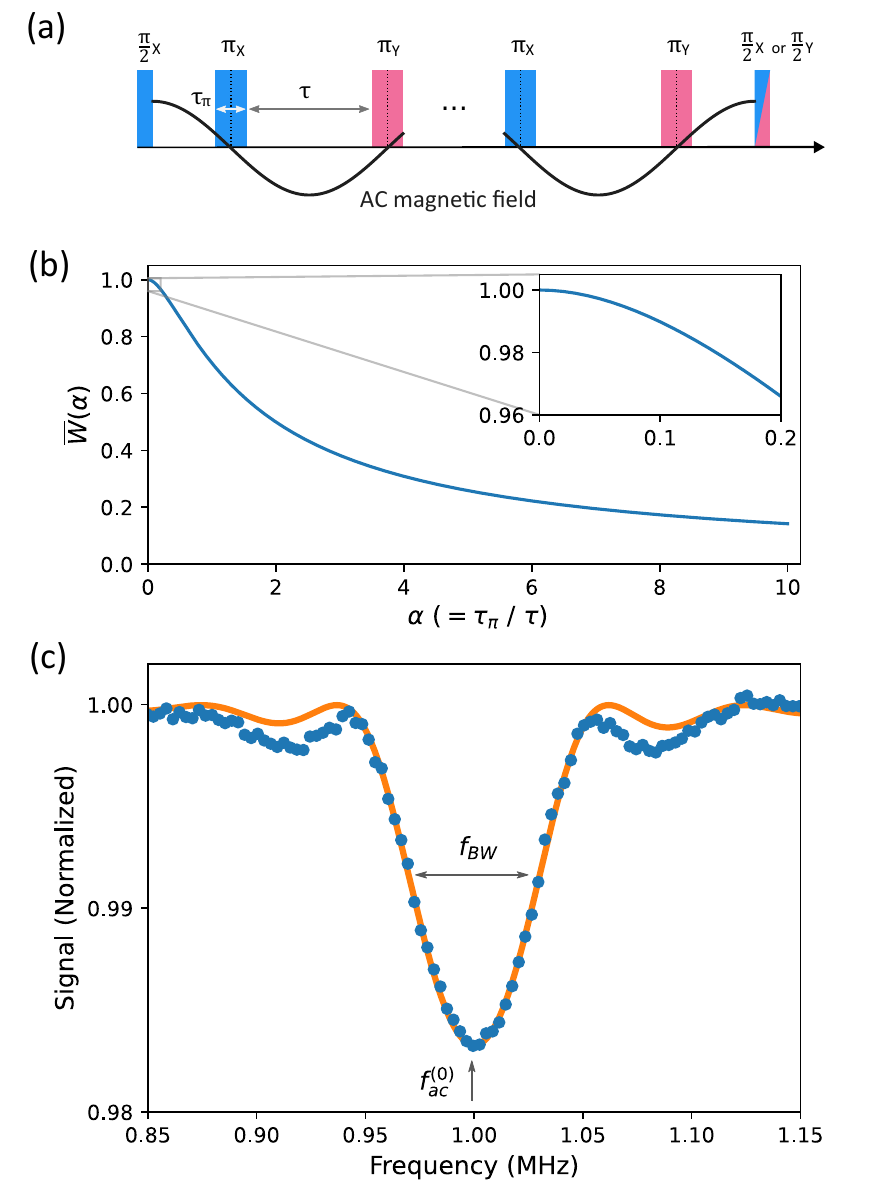}
	\caption{(a) The sequence of XY4-(k) in the presence of an AC magnetic field commensurate with the $\pi$ pulses. (b) The normalized weighing function $\overline{W}$ (see the main text) is plotted as a function of $\alpha$ (=$\tau_{\pi} / \tau).$ (c) The bandwidth curve of the AC sensing employing XY4-(4) is shown at the center frequency ($f_{ac}^{(0)}$) of 1 MHz. The theoretical curve fitted with only a single parameter to match the dip height is shown in orange.
 \label{fig1}}
\end{figure}

\section{Results and discussion}

\subsection{The effect of finite-width $\pi$ pulses}
In the following two subsections, we provide readers with a concise reminder of how the finite width of $\pi$ pulses influences the conditions of AC sensing protocols. Furthermore, we highlight the key features of the SR method, which will be exploited to interpret the frequency dependance of the sensitivity.
	For an oscillating field $B(t) = B_{ac} \cos( 2 \pi f_{ac}t + \phi _{ac})$ to be commensurate with the $\pi$ pulses, as illustrated in Fig. 1 (a), the frequency $f_{ac}$ should satisfy the condition below\cite{Degen2017, Ishikawa2018}
	\begin{equation}
		f_{ac} = \frac{1}{2\tau(1 + \alpha)},
		\label{f_ac}
	\end{equation}
	where $\alpha = \tau_{\pi}/\tau$ given $\tau$ and $\tau_{\pi}$ as illustrated in Fig. 1(a).  For Carl-Purcell type sequences, the accumulated phase $\Phi$ can be written as $\Phi$ = $\gamma B_{ac} N \tau (1 + \alpha) W$, in which $N$ is the total number of the $\pi$ pulses. According to Ref.~\cite{Ishikawa2018}, the weighting function $W$ is given as
	\begin{equation}
		W(\alpha, \beta, \phi_{ac}) = \frac{\sin(N \beta)}{N\beta} \left[ 1 - \frac{\cos(\beta \frac{\alpha}{1+\alpha})}{\cos(\beta)}\right]\cos(N\beta + \phi_{ac}),
		\label{W}
	\end{equation}
	where $\beta$ is defined as $\pi f_{ac} \tau (1 + \alpha)$. If the frequency $f_{ac}$ satisfies the condition of Eq.~(\ref{f_ac}), $\beta = \pi/2$. Then, $\alpha$ and the phase offset $\phi_{ac}$ crucially affect the accumulated phase $\Phi$. In Fig. 1(b), the normalized weighting function $\overline{W}(\alpha)$ (= $\frac{\pi}{2} W(\alpha, \phi_{ac}=0)$) is depicted as a function of $\alpha$. If $\alpha < 0.2$, the reduction of $\overline{W}(\alpha)$ is less than 4 $\%$. As $\alpha$ further increases, $\overline{W}(\alpha)$ shows a significant decrease, which implies that the DD sequence becomes highly insensitive to AC magnetic fields. In  Eq.~(\ref{W}), $\frac{\alpha}{1+\alpha}$ can be substituted by $\frac{f_{ac}}{f_{\mathrm{Rabi}}}$, in which Rabi frequency $f_{\mathrm{Rabi}}$ is defined as $\frac{1}{2 \tau_{\pi}}$. As $f_{ac}$ approaches to $f_{\mathrm{Rabi}}$, $\alpha$ becomes infinitely high. This explains that the AC-field sensing protocol is not able to detect the frequencies higher than Rabi frequency ($f_{\mathrm{Rabi}}$).

\begin{figure}[b!]
	\includegraphics[width=\columnwidth]{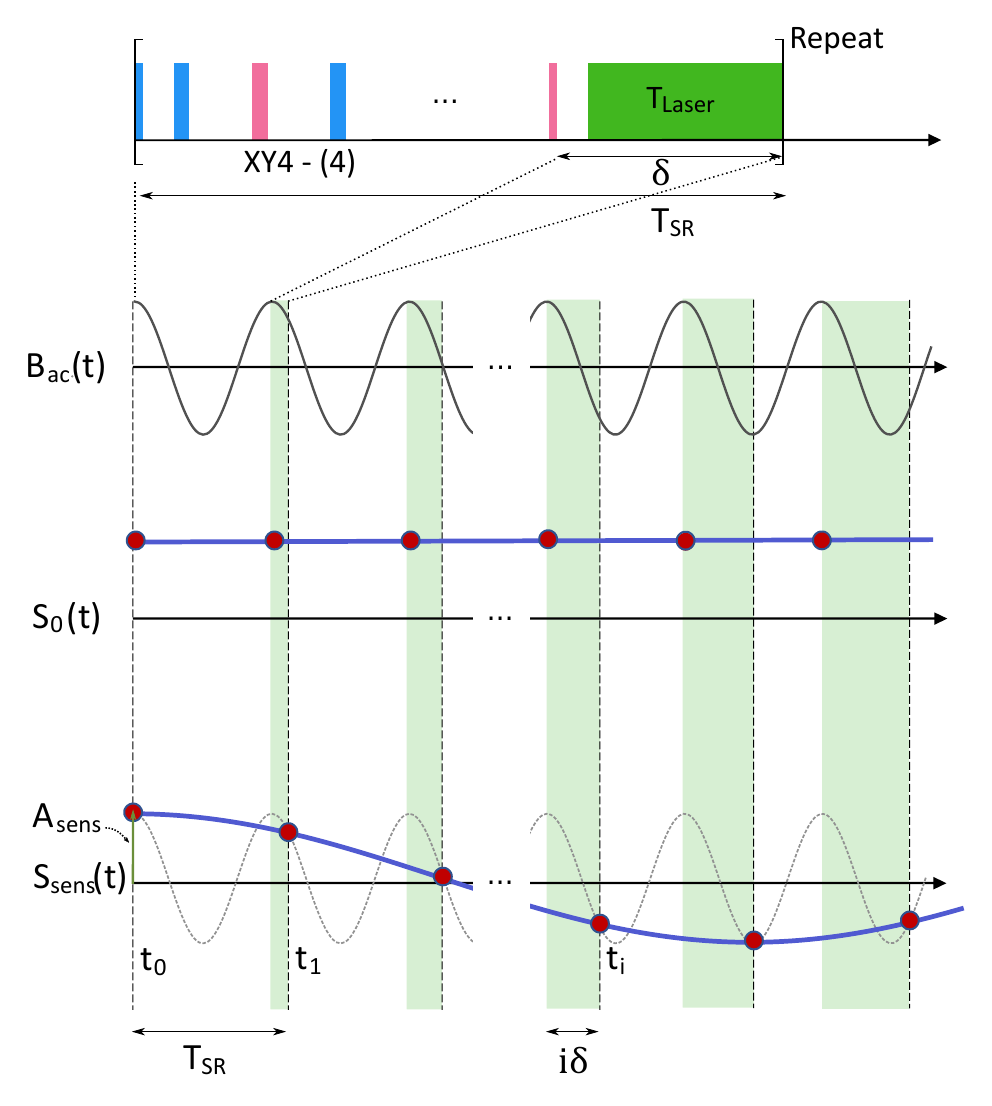}
	\caption{The principle of sequential readout is illustrated. The top exhibits the unit block and its duration is $\mathrm{T_{SR}}$. The unit block includes DD (XY4-(4)) sequence and optical readout/repolarization.  If $\mathrm{T_{SR}}$ is commensurate with the period of $B_{ac} (t)$, a constant signal S$_0$ (t) will be observed. If not, an under-sampled signal S$_{\mathrm{sens}}$ (t) is acquired and its frequency is given by Eq.~(\ref{f_sens}). }
		\label{fig2}
\end{figure}

\subsection{Sequential readout method}
	The principle of the SR method is illustrated in Fig.2. The unit block of SR consists of a DD sequence for sensing AC fields and laser pulse for optical readout and repolarization. The total duration $\mathrm{T_{SR}}$ splits into the interrogation time, $\tau_{sens} = N \tau (1 + \alpha)$, and the rest $\delta$. The duration of $\delta$ includes the laser pulse, $\mathrm{T_{Laser}}$ and a time margin that is practically inevitable. In general $\delta$ may not be commensurate with $B_{ac}(t)$ as the case in Fig.2. According to Eq.~(\ref{W}), the weighting function depends on the phase $\phi_{ac}$ of the ac field $B_{ac}(t)$. At the $i$th sampling time t$_i$ (=$i \mathrm{T_{SR}}$), the starting phase can be expressed as $2 \pi f_{ac} i \delta$, which is illustrated by the increasing areas in light orange color in Fig.~2. Because  the SR method entails an undersampling of $B_{ac}(t)$, the obtained signal S$_{sens}(t)$ has a down-converted frequency $f_{sens}$ given as
	\begin{equation}
		f_{sens} = f_{ac} - \overline{n}  f_{\mathrm{SR}},
		\label{f_sens}
	\end{equation}
	in which the integer $\overline{n} = \mathrm{argmin}_n\left( |f_{ac} - n f_\mathrm{SR}|\right)$, and the sampling frequency $f_{\mathrm{SR}} = 1/\mathrm{T_{SR}}$ (see the supplementary material for derivation). Eq.~(\ref{f_sens}) implies that the measured frequency $f_{sens}$ is determined by the sampling frequency $\mathrm{f_{SR}}$, irrespective of DD sequence. But, the frequency $f_{ac}$ should be within the sensible bandwidth ($f_{\mathrm{BW}}$ in Fig. 1 (c)). Then, the amplitude of the measured signal,  $A_{sens}$, can be expressed as
	\begin{equation}
		A_{sens} = S \cdot C \cdot \sin\left(\Phi_{(\phi_{ac}=0)}\right).
		\label{Amp}
	\end{equation}
	$S$ is the intensity of the optical signal acquired during $\mathrm{T_{Laser}}$, and $C$ is the contrast induced by spin-manipulating pulses within the DD sequence. The term of $\sin(\Phi)$ appears when the readout pulse of the DD sequence is $\frac{\pi}{2}(Y)$ as shown in Fig.2. For a weak AC magnetic field, $\sin(\Phi)$ can be approximated as $\Phi$.

\begin{figure}[t!]
	\includegraphics[width=1.05\columnwidth]{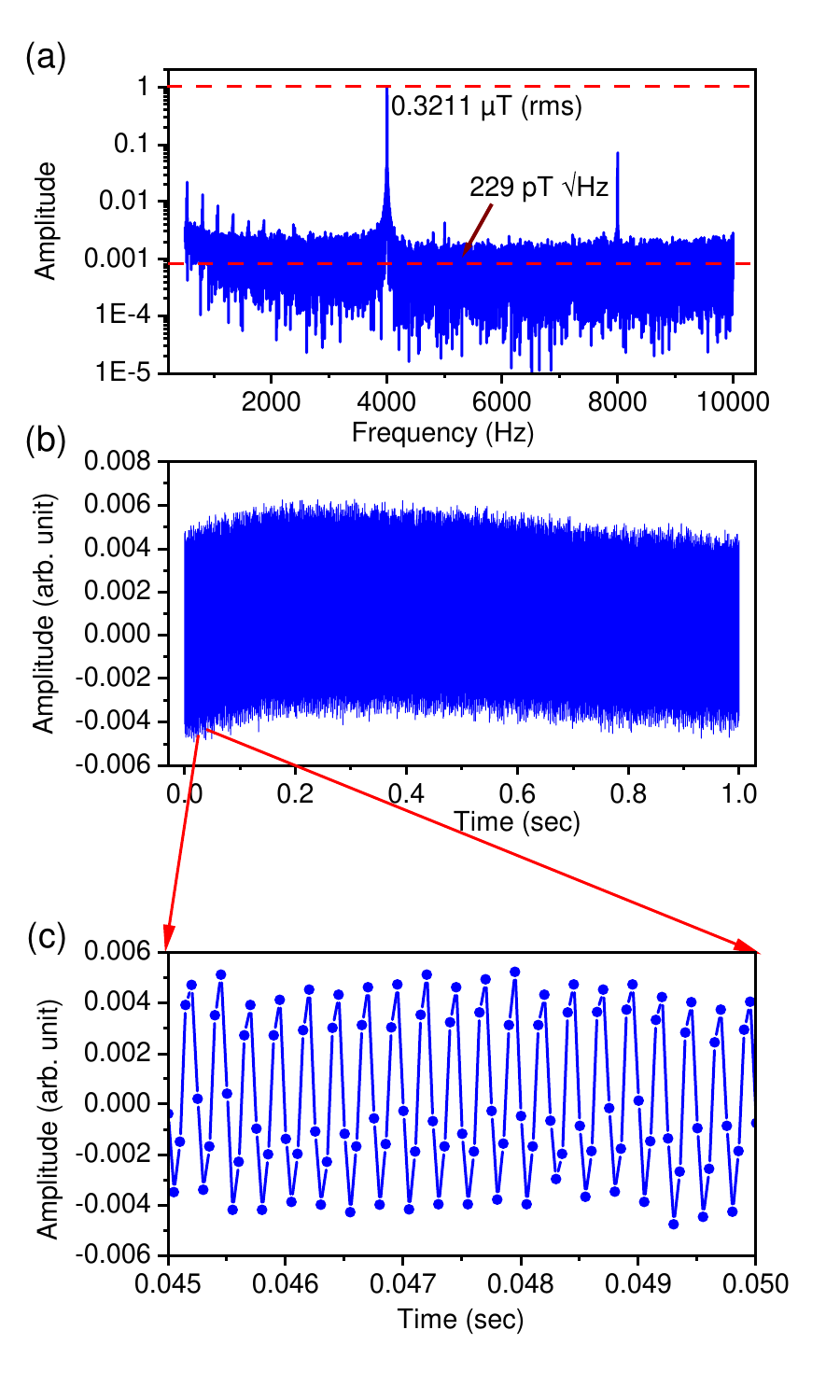}
	\caption{(a) Fourier transform of the time-domain signal S$_{\mathrm{sens}}$ (t) obtained by the SR method. The rms noise floor is calibrated from the intensity of the applied AC field, 0.3211 $\mu$T (rms), and the sensitivity is 229 pT/$\sqrt{\mathrm{Hz}}$. (b) The time-domain data of the oscillating signal of NV fluorescence measured for 1 s, and (c) its zoomed view for 5 ms.
		\label{fig3}}
\end{figure}

\subsection{Bandwidth}
	We experimentally determined the key features of NV AC-magnetometry using the XY4-(4) sequence. We choose this sequence because it exhibited the highest slope in the variation of the contrast as a function of AC-field amplitude (see the section 4 of the supplementary material). Since the readout pulse is $\frac{\pi}{2}(X)$, the output signal is proportional to $\cos(\Phi)$, and the formation of a dip happens when the frequency of external AC field satisfies Eq.~(\ref{f_ac}). In the presence of 1 MHz AC field with unknown intensities, we scanned $\tau$ and found the dips at theoretically expected values (see the section 3 of the supplementary material). For the main peak, $\tau$ should be 420 ns because $\tau_{\pi}$ = 80 ns in our experiment. With such $\tau$ and $\tau_{\pi}$ ($\alpha = 0.19$),  the intensity of the external 1MHz AC field was calibrated. As the amplitude $B_{ac}$ of the AC field increases,  the contrast of NV magnetometry follows a periodic oscillation as $\cos(\Phi)$ curve. The first minimum point corresponds to the phase accumulation of $\pi$. From the equation of $\Phi = \pi$, the amplitude $B_{ac}( \pi)$ of AC magnetic signal for the $\pi$ phase accumulation can be obtained as below
	\begin{equation}
		B_{ac}( \pi) =	\frac{ \pi^2 f_{ac}}{\gamma N \overline{W}(\alpha)}.
	\end{equation}
	The numerically calculated value of $\overline{W}(0.19)$ (= 0.968) was used for our calibration.
	
	In order to measure the bandwidth of the XY4-(4) sequence, we swept the frequency of AC field across 1 MHz. The obtained data was compared with the theoretical prediction $\cos(\Phi(f_{ac}))$, where $\Phi(f_{ac}) = \frac{\gamma N B_{ac}}{\pi f_{ac}^{(0)}} \overline{W}(f_{ac})$. In Eq.~(\ref{W}), $\beta = \frac{\pi}{2} f_{ac} / f_{ac}^{(0)}$, $f_{ac}^{(0)}$ = 1 MHz, $N$ = 16, and $\phi_{ac}$ = 0. The intensity of an ac field we applied is 1.9 $\mu$T. The orange line in Fig.1 (c) is the result of the fitting, where we only used a single parameter to match the height of the dip. The obtained data shows a good agreement with the theoretical prediction (orange), and the bandwidth $f_{BW}$ is found out to be approximately 50 kHz. 
	
\subsection{Sensitivity}
\begin{figure}[t!]
	\includegraphics[width=\columnwidth]{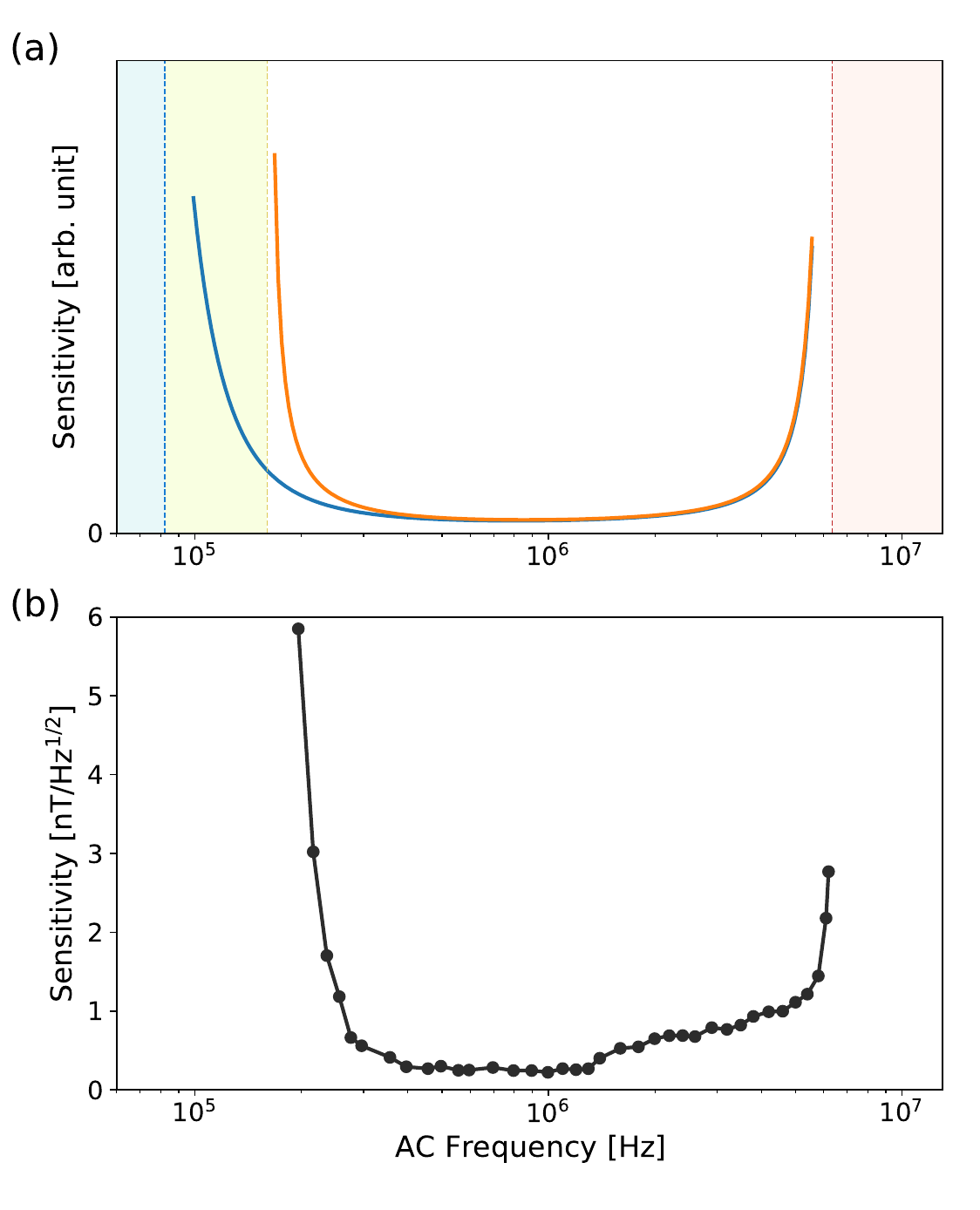}
	\caption{The frequency dependence of AC sensitivity based on the SR method is investigated. (a) Theoretical prediction according to Eq.~(\ref{eta}) is plotted in orange color. The red dashed line on the left side represents Rabi frequency limit (6.3 MHz). The lower limit is shown in orange dashed line (160 kHz). The blue dashed line (82 kHz) indicates T$_2$ limit, which is obtained by omitting $S(\mathrm{T_{Laser}})$ from Eq.~(\ref{eta}). (b) Experimentally obtained AC sensitivity as a function of frequency is in accordance with theoretical prediction.
		\label{fig4}}
\end{figure}

	To measure the sensitivity of AC magnetic fields, we used the SR method, of which main principle is described above. For a period of 1 s, we acquired the oscillating signal, $S_{sens}(t)$. In the presence of a reference AC-field having a calibrated rms intensity,  the sensitivity can be estimated from the signal to noise ratio (SNR) after Fourier transformation. In the experiment, we used the XY4-(4) sequence to measure the sensitivity. $\mathrm{T_{SR}}$ is set to 50 $\mu$s and repeated 20,000 times. The $\tau$ value is adjusted for 1 MHz.  But, the actual frequency of the reference AC-field was slightly detuned by 4 kHz ($f_{ac}$ = 996 kHz) to avoid being commensurate with the sampling rate $f_\mathrm{SR}$.  The detuning causes insignificant deterioration of the obtained signal, because 4 kHz is far less than the bandwidth f$_\mathrm{BW}$ ( = 50 KHz). The noise floor, which is indicated by the red dashed line below in Fig. 3(a), can be converted to detectable magnetic field $\delta B$  by multiplying the ratio between the rms intensity of the applied magnetic field (0.3211 $\mu$T) and the peak intensity at 4 kHz in the Fourier transformed spectrum (0.3211 $\mu$T corresponds to 34 mV$\mathrm{_{PP}}$ shown in Fig. S4 in the supplementary material). Because the total measurement time corresponds to 1 s, the sensitivity $\eta$ can be obtained as $\eta$ = $\delta B$/$\sqrt{\mathrm{Hz}}$. As shown in Fig. 3 (a), we obtained the sensitivity of 229 pT/$\sqrt{\mathrm{Hz}}$. The main peak appears at 4 kHz, which is consistent with Eq.~(\ref{f_sens}). The 2nd harmonic peak is observed due to the non-linearity in the response to the amplitude of AC field (refer to Fig. S4 in the supplementary material). Fig.3 (b) shows the time series data of oscillating signal of NV fluorescence measure for 1 s, and Fig. 3(c) is its zoomed view for 5 ms.

\subsection{Frequency dependence of sensitivity}
	We measured the frequency dependence of the sensitivity obtained by the SR method. If the dwell time $\mathrm{T_{SR}}$ is fixed, the sensitivity in Fig.~3 will be proportional to the amplitude $A_{sens}$ of the time-domain signal in Eq.~(\ref{Amp}). In varying the sensing frequency,  $\tau$ should be adjusted to meet the condition of Eq.~(\ref{f_ac}) for each $f_{ac}$ and thereby $\beta = \pi/2$ in Eq.~(\ref{W}). For producing AC magnetic fields in experiment, we used a single turn coil, of inductance is not negligible. Knowing the AC field intensity at 1 MHz attained from the calibration procedure in the sec. 2.3, we obtained the actual AC field intensities at other frequencies by comparing the current values (see the current variation in the supplementary material). For theoretical prediction, the parameters $S$, $C$, and $\Phi$ in Eq.~(\ref{Amp}) need to be expressed in terms of $f_{ac}$. $\overline{W}$ is relatively simple because $\frac{\alpha}{1+\alpha} = \frac{f_{ac}}{f_{\mathrm{Rabi}}}$ in Eq.~(\ref{W}) and $\phi_{ac}$ can be set $0$. The contrast $C$ in Eq.~(\ref{Amp}) decays as a function of $\tau$ due to decoherence process like $C(\tau/T_2)   =  C_0 \exp(-(N \tau/T_2)^p)$, in which $p$ and $T_2$ are estimated from decoherence curve fitting (see the supplementary material for details). $\tau$ varies with $f_{ac}$  following Eq.~(\ref{f_ac}) as $\tau$ = $1/2( 1/f_{ac} - 1/f_{\mathrm{Rabi}})$. Fig. 5(a) shows that the signal intensity $S$ increases with $\mathrm{T_{Laser}}$, because the degree of NV spin polarization to $m_S = 0$ state becomes higher. This repolarization process can be expressed as $S(\mathrm{T_{Laser}}/T_{p})  =  S_0 ( 1 - \exp(-\mathrm{T_{Laser}}/T_{p}))$ with the polarization time constant $T_{p}$. The numerical fitting found $T_p$ to be 15 $\mu s$. Finally, $\mathrm{T_{Laser}}$ is related with $f_{ac}$ through $\mathrm{T_{SR}}$. The relationship is described by $\mathrm{T_{Laser}}$ = $\mathrm{T_{SR}}$ - $N\tau (1+\alpha)$ = $\mathrm{T_{SR}}$ - $N /2 f_{ac}$. Including them altogether, we can describe how the sensitivity varies as a function of $f_{ac}$,
	\begin{equation}
		\eta(f_{ac}) \propto \frac{1}{S(\mathrm{T_{Laser}}/T_{p}) C(\tau/T_2) \Phi(f_{ac})} \sqrt{\frac{\mathrm{T_{SR}}}{N \tau(1 + \alpha)}}.
		\label{eta}
	\end{equation}

The first terms originate from the inverse of the amplitude in Eq.~(\ref{Amp}), and the last term represent the effect of the overhead time.\cite{Pham2012,Edlyn2019} The theoretical prediction is shown as orange curve in Fig. 4 (a). The experimental data in Fig.4 (b) shows a good agreement with it. The high-frequency limit is clearly determined by  the Rabi frequency, being approximately 6.3 MHz in the present study, which is indicated by the red dashed line. The low-frequency limit is close to 200 kHz (yellow dashed line). We found that the effect of decoherence causing the decay of the contrast $C(\tau/T_2)$ alone cannot explain it. The blue curve in Fig. 4 (a) is obtained by omitting $S(\mathrm{T_{Laser}}/T_{p})$ from Eq.~(\ref{eta}) and its low-frequency limit is below 100 kHz, indicated by blue dashed line. This discrepancy originates from the reduction of optical signal intensity $S$ when the duration of T$_{\mathrm{Laser}}$ decreases (Fig. 5 (a)). Given T$_{\mathrm{SR}}$, lowering $f_{ac}$ accompanies decreasing T$_{\mathrm{Laser}}$ as seen by the fitting equation $\mathrm{T_{Laser}}$ = $\mathrm{T_{SR}}$ - $2N / f_{ac}$ in Fig.~ 5 (b). The reduced signal intensity results in the deterioration of the sensitivity. In addition, similar to previous works\cite{Pham2012,Edlyn2019}, the theoretical prediction described by Eq.~(\ref{eta}) has the optimal frequency, which is found to be within the range from 800 kHz to 1 MHz. We achieved a sensitivity of 229 pT/$\sqrt{\mathrm{Hz}}$ at 1 MHz (Fig. 3(a)).

\begin{figure}[t!]
	\includegraphics[width=\columnwidth]{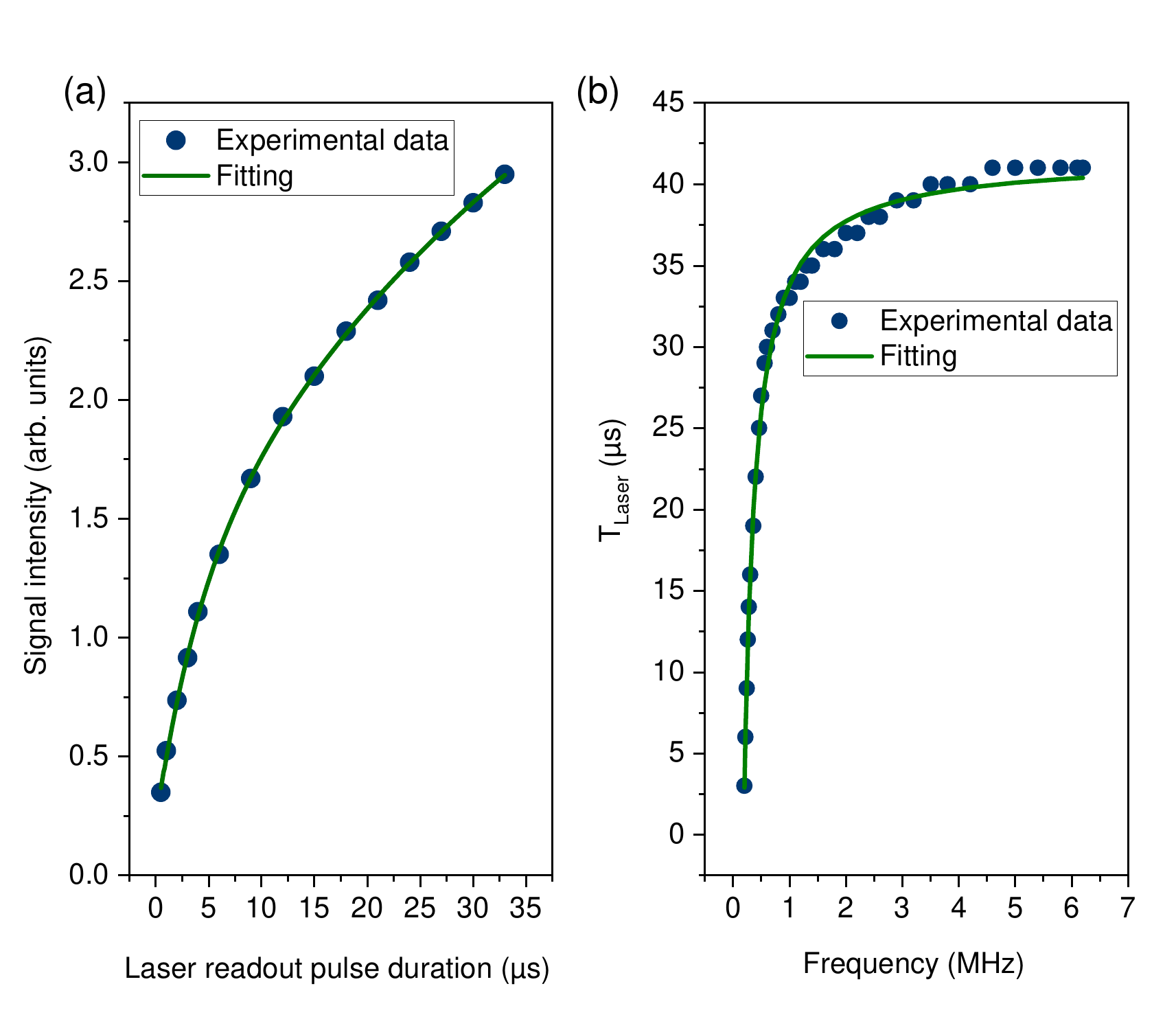}
	\caption{(a) NV fluorescence intensity is measured as a function of optical readout duration. The repolarization time $t_p$ is estimated to be 15 $\mu$s from exponential fitting. (b) The duration  T$_{\mathrm{Laser}}$ of optical readout and repolarization varies according to the sensing frequency because the dwell time T$_{\mathrm{SR}}$ is fixed.  The values of T$_{\mathrm{Laser}}$ used in a series of measurement nearly follow the relation, $\mathrm{T_{Laser}}$ = $\mathrm{T_{SR}}$ - $N /2 f_{ac}$.
		\label{fig5}}
\end{figure}

In Eq.~(\ref{eta}), the term $S(\mathrm{T_{Laser}}/T_{p})$ sets it apart from earlier theoretical predictions\cite{Pham2012,Edlyn2019}. Governed by the ratio $\mathrm{T_{Laser}}/T_{p}$, mitigating the impact of $S(\mathrm{T_{Laser}}/T_{p})$ and thus reaching  the lower frequency limit determined by $\mathrm{T_2}$ can be accomplished through two approaches. The first method is associated with the optical power density irradiated to NV centers. The power density for optical readout/repolarization should be high enough to sufficiently polarize NV spins within $\mathrm{T_{Laser}}$. With a laser spot diameter of 0.2 mm and laser power of 1.4 W in this work, the power density utilized is approximately 45 W/mm$^2$. However, this value still falls significantly short of the saturation density of NV spins ($>$ 5 kW/mm$^2$) as estimated from single NV experiments.\cite{Siyushev2019} Given the considerably high saturation power density for NV spins, practical challenges might arise in achieving complete optical repolarization during the $\mathrm{T_{Laser}}$ time with an ensemble of NV centers.  The second method involves increasing $\mathrm{T_{Laser}}$. This leads to the increase in the dwell time $\mathrm{T_{SR}}$ of the SR method. Essentially, the SR method emulates the acquisition of a highly under-sampled signal as shown in Fig. (2). If a sinusoidal signal in the time domain is subjected to a certain amount of Gaussian noise, the noise floor of the under-sampled signal decreases at a higher sampling rate. This effect stems from the fact that higher sampling rates distribute the noise across a wider frequency. Hence, during the measurements for Fig. 4 (b), we maintained a fixed value for $\mathrm{T_{SR}}$ to prevent variations in the noise floor due to changes in bandwidth. When employing  sequential readout, the selection of $\mathrm{T_{SR}}$ is imperative in light of the desired bandwidth. What Eq.~(\ref{eta}) implies is that the consideration of $\mathrm{T_{Laser}}$ and $T_p$ plays an  essential role in anticipating the lower frequency limit.

\section{Conclusion}
In this work, we present a combination of experimental and theoretical investigations into the frequency dependent behavior of AC magnetic field sensitivity through the use of sequential readout. Our findings unveil that the upper limit is determined by the finite-width of $\pi$ pulses, while the lower limit can be governed by the duration of optical repolarization for NV ensembles when the optical power density is often below the saturation limit. With the dwell time of the SR method held constant, our theoretical model, which encompasses these two factors, demonstrates a reasonable concordance with the experiment results. Furthermore, through the employment of the XY4-(4) DD sequence, we obtained a maximum sensitivity of 229 pT/$\sqrt{\mathrm{Hz}}$ at 1 MHz.   Considering the sensitivity reported in the previous study\cite{Glenn2018}, ranging from 30 to 70 pT/$\sqrt{\mathrm{Hz}}$, we hold the conviction that the achieved sensitivity empowers us to conduct NV-NMR in micron-scale, where (thermal) Boltzmann polarization remains predominant.\cite{Allert2022} This work not only offers valuable insight but also furnishes practical guidelines for the applications of AC magnetometry using NV spins in the range of kilohertz to megahertz. These applications may include other areas, such as magnetic induction tomography, in which NV spins are expected to facilitate high spatial resolution imaging.

\section*{acknowledgments}
This research was supported by a grant (GP2022-0010) from Korea Research Institute of Standards and Science, and Institute of Information and communications Technology Planning $\&$ Evaluation (IITP) grant funded by the Korea government (MSIT) (No.2019-000296).
\\
\appendix
\section{Experimental methods}
\subsection{optical and microwave system}
A static magnetic field of 5.4 mT is aligned along one of the four NV orientations in diamond. A continuous-wave laser (532 nm) is used to excite the NV spins. The laser passes through an acousto-optic modulator (AOM). An arbitrary waveform generator is used to store the dynamical decoupling (DD) sequences, and a signal generator to produce the microwave frequency. After mixing these two signals by frequency mixer, 100 watt amplifier amplifies the signal and delivered to the diamond sample through 2 mm microwave loop  below the sample.  Timing signals are generated from a programmable timing generator and control the outputs of AWG, AOM and microwave through the RF switches. The fluorescence from the diamond NV centers is collected with a parabolic concentrator and delivered to the photodiode. A long pass filter is placed between the concentrator and photodiode, allowing only NV fluorescence to pass through. The collected signal from the photodiode is amplified by a current amplifier and then recorded by digital-to-analog converter. Further details of experimental method are described in the supplementary material.

\subsection{Diamond sensor}
A thin nitrogen-doped ([$^{14}$N] $\simeq$ 10 ppm) diamond layer ($^{12}$C $>$ 99.99$\%$, 40 $\mu$m thick) is grown by chemical vapor deposition (CVD) on top of an electronic grade diamond plate by Applied Diamond. The dimensions of the diamond plate are approximately 2 × 2 × 0.54 mm$^3$ . The diamond is electron irradiated (1 MeV, 1 × 10$^{19}$/cm$^2$ ), and annealed in a vacuum at 800 $^{\circ}$C for 4 hours and 1000 $^{\circ}$C for 2 hours, sequentially.


%

\end{document}